# Superbackscattering Nanoparticle Dimers

Iñigo Liberal,[1] Iñigo Ederra,[1] Ramón Gonzalo,[1] and Richard W. Ziolkowski[2]

[1]*Electrical and Electronic Engineering Department,*
*Universidad Pública de Navarra, Campus Arrosadía, Pamplona, 31006 Spain*
[2]*Department of Electrical and Computer Engineering, University of Arizona, Tucson, AZ, 85721 USA*

The theory and design of superbackscattering nanoparticle dimers are presented. We analytically derive the optimal configurations and the upper bound of their backscattering cross-sections. In particular, it is demonstrated that electrically small nanoparticle dimers can enhance the backscattering by a factor of 6.25 with respect to single dipolar particles. We demonstrate that optimal designs approaching this theoretical limit can be found by using a simple circuit model. The study of practical implementations based on plasmonic and high-permittivity particles reveal that fourfold enhancement factors might be attainable even with realistic losses.

## I. INTRODUCTION

The scattering of electromagnetic fields by subwavelength particles is of fundamental interest for a wide range of disciplines including physics, optics and engineering [1–5]. In most cases, the response of electrically small particles can be approximated by the excitation of an electric dipole, usually aligned with the incident electric field, and whose value is directly proportional to the local electric field and particle volume [1, 2]. Despite this fact, researchers have aggressively pursued the excitation of more exotic and sophisticated scattering responses. These include, among others, magnetic dipoles [6–11], higher order modes (HOMs) [12–16], as well as their interdependence via magnetoelectric coupling [17, 18]. When properly combined, the excitation of these more advanced responses enables unusually large scattering behaviors, including the so-called superscattering [19, 20], and superdirective scattering [21–23]. Additional degrees of freedom can be introduced by using nonlinear elements [24]. On the other hand, the total [25–27] or angularly selective [28, 29] suppression of scattering is equally important, as well as the scattering minimization while maintaining a useful amount of absorbed power [30–32]. In addition to the research on specific configurations, there have also been substantial efforts in understanding the fundamental limits and physical bounds of both broadband [33–36] and time-harmonic [37–40] scattering by passive particles. Even if the limits of passive particles are too stringent, the scattering can be boosted by using active media, resulting in the development of subwavelength nanoparticle lasers [23, 41, 42].

Despite all of this active research in the field of scattering physics and engineering, little to no attention has been paid to *enhancing* the backscattering of subwavelength particles (i.e., to the reflection from, or to the power re-radiated against the direction of propagation of the incident wave). This fact is surprising in view of the broad scope of particles with large backscattering cross-sections. In essence, the operation of any spectroscopy, communication, remote sensing (radar, sonar, ...), manipulation and/or imaging system that makes use of a single emitter/receiver device relies on a backscattering measurement. This wide range of applicability should undoubtedly motivate the research on particles featuring enhanced backscattering cross-sections.

In a recent work, the authors have derived an upper bound for the time-harmonic backscattering cross-section (normalized to the wavelength squared) of a passive particle [37]

$$\sigma_b\left(-\widehat{\mathbf{k}}_i\right) \leq \frac{1}{\pi} D_{\text{scat}}\left(-\widehat{\mathbf{k}}_i\right) D_{\text{scat}}\left(\widehat{\mathbf{k}}_i\right) \qquad (1)$$

This bound illustrates the limit on backscattering as a function of the scattering directivity in the forward and backward directions, $D_{\text{scat}}(\widehat{\mathbf{k}}_i)$ and $D_{\text{scat}}(-\widehat{\mathbf{k}}_i)$, respectively (i.e., $\widehat{\mathbf{k}}_i$ stands for the direction of the incident wave). Intuitively, in order to maximize the scattering along a given direction, in this case the backward direction, the particle must not only focus its scattering along this direction, but also along the incident direction. This backward-forward requirement is necessary in order to produce the destructive interference required to extract energy from the incident field. Following this philosophy, one can fully align the design of superbackscattering particles with the superdirective beamforming methods developed in antenna engineering. In fact, this concept has already been successfully applied to the design of superbackscattering antenna arrays at microwave frequencies [43]. Following a similar philosophy, the present work investigates the design of superbackscattering nanoparticle dimers. These geometries have been selected in view of previous positive experiences with two-element superdirective antenna arrays [44–47] and plasmonic dimers [48].

## II. THEORY

Let us then begin by examining a nanoparticle dimer composed of two nanoparticles symmetrically located with respect to the origin on the z-axis with inter-particle separation $d$ (c.f., figure 1(a)). The dimer is illuminated by a time-harmonic ($e^{j\omega t}$ time dependence assumed throughout) $\widehat{\mathbf{x}}$-linearly polarized plane-wave propagating along the z-axis, i.e., $\mathbf{E}^i = \widehat{\mathbf{x}}\, E_0 \exp(-j\, k_0\, z)$. For subwavelength particles whose separation is larger than the particle radius, i.e., $d - (r_1 + r_2) > r_1, r_2$, the fields scattered by the nanoparticles can be fairly approximated by those radiated by an electric dipole $\mathbf{p}$, or, equivalently, an electric Hertzian dipole (EHD) with current moment $Id\mathbf{l} = \mathbf{p}/j\omega$. Under this dipolar approximation, the particle dimer can be treated as a two-element linear array of EHDs. In the far zone, the field scattered by the dimer consists of a spherical wave with electric field $\mathbf{E}^s = E_0 \mathbf{f}(\hat{\mathbf{r}}) \exp(-j k_0 r)/(k_0 r)$, where $\mathbf{f}(\hat{\mathbf{r}})$ is the (unit-less) scattering pattern of the dimer, given by

$$\mathbf{f}(\hat{\mathbf{r}}) = \mathbf{f}_{\text{dp}}(\hat{\mathbf{r}}) \left( A_1 e^{-j\frac{k_0 d}{2}\hat{\mathbf{r}}\cdot\hat{\mathbf{z}}} + A_2 e^{j\frac{k_0 d}{2}\hat{\mathbf{r}}\cdot\hat{\mathbf{z}}} \right) \quad (2)$$

where $\mathbf{f}_{\text{dp}}$ is the dipolar scattering pattern of each individual particle, and $A_1$ and $A_2$ are the excitations coefficients describing the relative magnitude and phase of the dipole excited at each nanoparticle. Consequently, the scattering directivity of the nanoparticle dimer can be written as follows

$$D_{\text{scat}}(\hat{\mathbf{r}}) = \frac{4\pi |\mathbf{f}(\hat{\mathbf{r}})|^2}{\int_0^{2\pi}\int_0^\pi |\mathbf{f}(\hat{\mathbf{r}}')|^2 \sin\theta' d\theta' d\phi'} \quad (3)$$

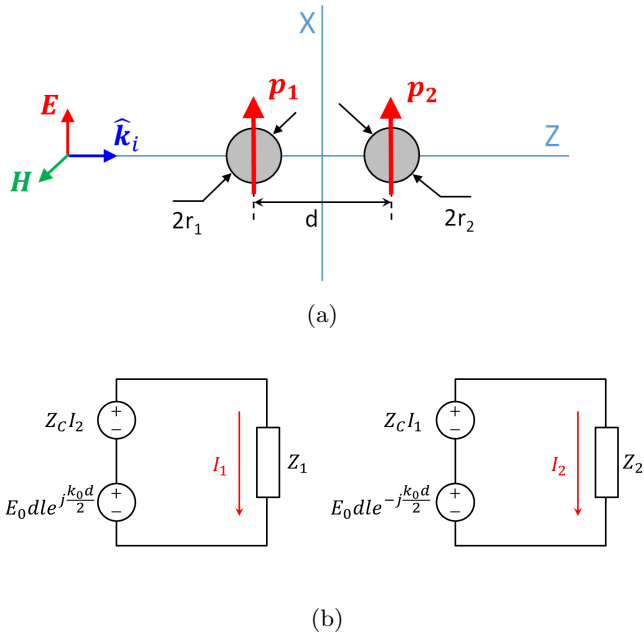

FIG. 1. (a) Sketch of the scattering geometry: a plane wave illuminates a nanoparticle dimer, exciting the electric dipole moments of each nanoparticle. (b) Equivalent circuit model.

In view of (1), the least upper bound of the backscattering cross-section of a nanoparticle dimer can be found by identifying the optimal excitation coefficients, $A_1 = A_1^{opt}$ and $A_2 = A_2^{opt}$, that maximize the forward-backward directivity product, $D_{\text{scat}}(-\hat{\mathbf{k}}_i) D_{\text{scat}}(\hat{\mathbf{k}}_i)$. In general, different excitation coefficients are obtained for each separation distance $d$, i.e., $A_p^{opt} = A_p^{opt}(d)$, for $p = 1, 2$. In particular, it is found in Appendix A that, in the limit of electrically small separation distances ($k_0 d \to 0$), the optimal excitation coefficients to maximize the backscattering are given by

$$A_2^{opt}(d) = -A_1^{opt}(d) = \frac{a_0}{k_0 d} \quad (4)$$

where $a_0 \in \mathbb{C}$ is an arbitrary complex constant.

Interestingly, the superbackscattering response is obtained exactly at $A_1^{opt} = -A_2^{opt}$, i.e., the electric dipole moments excited in each nanoparticle must be of the same magnitude but opposite direction. This simple condition has been previously mentioned in the literature under other circumstances, and it deserves a careful discussion. To begin, superdirective antenna articles relate opposing currents with radiation superdirectivity [44], which, in scattering terms, is directly associated with superdirective scattering in the forward direction. However, the maximum of the forward scattering directivity, also reported in the Appendix A, is obtained for slightly different coefficients $A_1^{fwd} = -a_0/(k_0 d)$ and $A_2^{fwd} = a_0 [1/(k_0 d) + j2/5]$. Note that the coefficients for optimal backscattering and optimal forward-scattering become very similar for small separation distances, i.e., $A_1^{fwd} \simeq A_1^{opt}$, $A_2^{fwd} \simeq A_2^{opt}$, despite the fact that they lead to significantly different patterns and scattering cross-sections. Moreover, the fact that significantly different responses are obtained with numerically similar coefficients emphasizes the stringent fabrication tolerances associated with the construction of superdirective devices [45].

It is also worth noticing that, when condition (4) is satisfied, the net electric dipole moment of the system: $\mathbf{p}_1 + \mathbf{p}_2$, is zero. This effect is usually associated with scattering cancellation configurations in many body problems [49]. Intriguingly, our analysis remarks that the backscattering is optimized when the predominant scattering response of the dimer is canceled out. Naturally, this is only possible if the usually secondary responses are strengthened. In particular, note that opposed electric currents have been traditionally associated with strong magnetic dipole moments [50]. Other works have also emphasized the role of the electric quadrupole excitation in these structures [51, 52]. To shed more light onto how the secondary responses are responsible for the enhanced backscattering, it is worth examining the multipolar decomposition of the above structure.

Without any loss of generality, the scattered field outside the particle can be written in multipolar form as follows:

$$\mathbf{E}^s(\mathbf{r}) = \sum_{\{q\}} \left[ a_{nm}^{lTM} \mathbf{N}_{nm}^{l>}(\mathbf{r}) + a_{nm}^{lTE} \mathbf{M}_{nm}^{l>}(\mathbf{r}) \right] \quad (5)$$

A complete description of the above notation can be found in [40]. For the configuration depicted in figure (1)(a), the only nonzero scattered field coefficients correspond to the set of even $TM$ modes, $TM^{en1}$, and odd $TE$, $TE^{on1}$, modes. These coefficients can be writ-

ten as follows

$$a_{n1}^{eTM} = [I_1 dl - (-1)^n I_2 dl] \left\{ 2(n+1)(k_0 d)^{n-1} \varphi_n \right\} \quad (6)$$

$$a_{n1}^{oTE} = [(-1)^n I_1 dl + I_2 dl] \left\{ (k_0 d)^n \varphi_n \right\} \quad (7)$$

with the term

$$\varphi_n = \frac{\eta_0 k_0^2}{4\pi} \frac{1}{2^n} \frac{1}{n(n+1)(2n-1)!!} \quad (8)$$

It is apparent from (6), (7) that the first dominant term is the electric dipole, which is $\propto (k_0 d)^0$. The next dominant terms are the magnetic dipole, $\propto k_0 d$, and electric quadrupole, $\propto k_0 d$. However, when $I_2 dl = -I_1 dl$, the dimer only excites the $TM^{en1}$ modes with $n$ being an even number, and the $TE^{on1}$ modes with $n$ being an odd number. Therefore, the net electric dipole ($TM^{e11}$ mode) is canceled out, and the radiation is then dominated by the magnetic dipole ($TE^{o11}$ mode) and electric quadrupole ($TM^{e21}$ mode) terms.

To finalize this analysis, let us inspect the maximum value of the backscattering cross-section. As derived in Appendix A, when the response of the individual particles can be approximated by their dipolar response, the upper bound of the backscattering cross-section of a nanoparticle dimer is given by

$$\sigma_b\left(-\hat{\mathbf{k}}_i\right) \leq \frac{1}{\pi}\left(\frac{3 \cdot 5}{4}\right)^2 \simeq 4.5 \quad (9)$$

Therefore, the upper bound (9) represents an enhancement by a factor 6.25 with respect to the bound of a single dipolar particle: $(1.5)^2/\pi \simeq 0.72$. Thus, in theory, nanoparticle dimers might present significant backscattering enhancements as compared to individual particles. However, as demonstrated in [37], the upper bound of the backscattering cross-section for a nanoparticle system in which a given number of multipoles: $N_p = 2$, are excited is

$$\sigma_b\left(-\hat{\mathbf{k}}_i\right) \leq \frac{1}{4\pi}\left(N_p^2 + 2N_p\right)^2 \simeq 5.1 \quad (10)$$

Consequently, in terms of backscattering, it is concluded that a nanoparticle dimer cannot fully exploit the potential of all of the dipole and quadrupole terms simultaneously, i.e., the aforementioned electric quadrupolar, $TM^{o21}$, and, magnetic dipole, $TE^{e11}$, modes cannot be optimally excited at the same time. This suggests that the scattering properties of coupled nanoparticle systems can be more intuitively assessed by using antenna array formulations, as was done with (2), instead of employing the entire multipolar decomposition. Moreover, one could then draw upon previous superdirective antenna results to guide the understanding of dimer superscattering behaviors.

### III. CIRCUIT MODEL

After identifying the magnitude and phase of the dipole moments that lead to a superbackscattering response, one must still find the correct design (nanoparticle geometry and materials) that provides the optimal excitation coefficients (4). We now demonstrate that this can be accomplished by using a simple circuit model. Specifically, the local field equations acting on each particle can be written in circuital terms as:

$$Z_1 I_1 = E_0 dl\, e^{j\frac{k_0 d}{2}} + Z_C I_2 \quad (11)$$

$$Z_2 I_2 = E_0 dl\, e^{-j\frac{k_0 d}{2}} + Z_C I_1 \quad (12)$$

This system of equations corresponds to the equivalent circuit model depicted in figure 1(b). In this circuit model, the source voltages are represented by the local electric field acting on each of the particles, integrated over the differential length of the current moment that reproduces its dipolar response, i.e., $I_1 dl = |\mathbf{p}_1|/j\omega$ and $I_2 dl = |\mathbf{p}_2|/j\omega$. Similarly, the current flowing in the circuit represents the magnitude of these current moments, $I_1$ and $I_2$. Therefore, $Z_1$ and $Z_2$ are the self-impedances associated with each nanoparticle current, and $Z_C$ is the coupling impedance, i.e., the usually written mutual impedances and their reciprocal relation $Z_{21} = Z_{12}$ are replaced by it. In general, the impedance term of each particle is given by $Z_p = -\eta_0(k_0 dl)^2/\left(6\pi\, b_{11}^{eTM}\right)$, where $b_{11}^{eTM}$ is the scattered field coefficient of the nanoparticle's electric dipole ($TM^{e11}$) mode [40].

This simple formulation serves to elucidate the response of each individual nanoparticle, their interaction, and, consequently, the nanoparticle dimer. For example, in lossless particles, the resistance reduces to the scattering resistance of the dipole mode: $R_1 = R_2 = R_{\mathrm{dp}} = \eta_0(k_0 dl)^2/(6\pi)$, which is independent of the nanoparticle's properties. In contrast, the reactance is strongly determined by the nanoparticle's characteristics. For example, electrically small dielectric particles with radius $r_1$ and relative permittivity $\varepsilon$ are characterized by the capacitive reactance $X = 1/(-\omega C)$, with $C = 4\pi\, \varepsilon_0\, \left(r_1^3/dl^2\right)\left[(\varepsilon - 1)/(\varepsilon + 2)\right]$. On the other hand, electrically small plasmonic nanospheres with relative permittivity described by the lossless Drude model: $\varepsilon = 1 - \omega_p^2/\omega^2$, are characterized by a series LC circuit whose reactance is: $X = \omega L + 1/(-\omega C)$, with $L = 3\left(dl^2/r_1^3\right)/(4\pi\, \varepsilon_0\, \omega_p^2)$, $\omega_p$ being the plasma frequency, and $C = 4\pi\, \varepsilon_0\, \left(r_1^3/dl^2\right)$.

Next, the local field acting on the nanoparticle is given by the addition of the incident electric field plus the scattered electric field produced by the neighboring nanoparticle. The latter is described via the coupling impedance, $Z_C$, which can be determined by inspecting the electric field excited by an EHD and dividing it by its current moment:

$$Z_C = \eta_0 \frac{(k_0 dl)^2}{4\pi}\left[\frac{-j}{k_0 d} + \frac{-1}{(k_0 d)^2} + \frac{j}{(k_0 d)^3}\right] e^{-jk_0 d} \quad (13)$$

For small separation distances, $Z_C$ can be approximated as:

$$Z_C = R_C + jX_C \approx -R_{\mathrm{dp}} + R_\delta + j\frac{1}{\omega C} \qquad (14)$$

Thus, the reactance of the coupling impedance is dominated by a negative capacitance (positive reactance), with $C \simeq 4\pi\,\varepsilon_0\,d^3/dl^2$. In particular, the dispersion profile of the coupling reactance is inverted with respect to that of the particle's self-impedance. This ensures that the structural (dimer) resonances only occur at discrete frequencies. Moreover, it also means that the resonance phenomena cannot extend over an unlimited bandwidth in passive particles. In addition, there is a small and negative resistance whose magnitude approximately equals the dipolar scattering resistance. In order to avoid singularities it is convenient to point out that the absolute value of this resistance is actually slightly smaller. This correction value can be approximated as $R_\delta \simeq R_{\mathrm{dp}}\,(k_0 d)^2/5 > 0$.

The solution to the system of equations (11)-(12) provides the current moments excited in the nanoparticles

$$I_1 = \frac{Z_2 + Z_C e^{-jk_0 d}}{Z_1 Z_2 - Z_C^2}\,E_0 dl\,e^{j\frac{k_0 d}{2}} \qquad (15)$$

$$I_2 = \frac{Z_1 + Z_C e^{jk_0 d}}{Z_1 Z_2 - Z_C^2}\,E_0 dl\,e^{-j\frac{k_0 d}{2}} \qquad (16)$$

It is apparent from (15), (16) that the resonance condition, in which strong currents are excited in both particles despite their electrically small size, is approximately given by $X_1 X_2 = X_C^2$. In addition, the reactances required to satisfy the condition $I_1 = -I_2$ are given by

$$X_1 = -X_C - \frac{R_\delta}{k_0 d} \qquad (17)$$

$$X_2 = -X_C + \frac{R_\delta}{k_0 d} \qquad (18)$$

It can be readily checked that the impedances (17), (18) match almost perfectly with the resonance condition $X_1 X_2 = X_C^2$. Therefore, it can be concluded that it is possible to excite very strong and opposing currents in coupled nanoparticle systems. Moreover, the circuit model formulation provides us with a simple way to design the superbackscattering nanoparticle dimers: First, it is evident from (17), (18) that the particles must feature a capacitive reactance, meaning that the superbackscattering resonance is shifted toward smaller frequencies (longer wavelengths) with respect to the particle resonance. Second, since $|X_C| \gg |R_\delta/k_0 d|$, it follows that nanoparticle dimers composed of identical particles actually would approximately fit to the superbakcscattering condition. Therefore, they can be considered as a good starting point in the design process. Finally, since $X_1 < X_2$ in (17), (18), the design can be refined further by increasing the resonance frequency of the first particle (or decreasing the resonant frequency of the second particle). We emphasize that this design process is quite general, and it can be applied to any dimer constituted by dipolar particles.

## IV. PLASMONIC NANOPARTICLE DIMERS

As a first example of the use of the proposed methodology, we study the implementation of dimers composed of plasmonic nanoparticles. Specifically, we assume that the plasmonic particles are made of aluminum-doped zinc oxide (Al:ZnO) semiconductors [53, 54], characterized by a relative permittivity that is described by a lossy Drude dispersion model: $\varepsilon = 1 - \omega_p^2/(\omega^2 - j\omega\omega_c)$, with plasma frequency $\omega_p = 2\pi\cdot 2213.2\,\mathrm{THz}$. For the case of simplicity, the lossless, $\omega_c = 0$, case is considered first, though realistic losses $\omega_c = 2\omega_p \times 10^{-3}$ will be addressed later. The relatively low-loss plasmonic response of Al:ZnO is particularly convenient for the design of nanoparticle dimers. In fact, subwavelength Al:ZnO nanoparticles exhibit strong dipole resonances at near-IR and short-wavelength IR frequencies. Another attractive aspect is that the plasma frequency of Al:ZnO (and hence the resonance frequency of the nanoparticles) can be tuned by adjusting the doping level [53].

The starting point of the design process consists of a dimer with identical particles of $r_0 = 100\,\mathrm{nm}$ radius, separated by the distance $d = 3\,r_0$. Following our circuit model approach, figure 2 depicts the reactance of the individual nanoparticles composing the dimer, $X_1 = X_2$, as well as the coupling reactance between them, $X_C$. As expected, the frequency dispersion of the reactance of the plasmonic nanoparticles approximates that of a series LC circuit (capacitive at low frequencies, while inductive at frequencies above the resonance, i.e., a change in sign from one side of the resonance to the other). The resonance frequency of each individual nanoparticle ($X_1 = X_2 \simeq 0$) takes place approximately at $\lambda = 1537\,\mathrm{nm}$, and it is denoted by the vertical black dotted line. On the other hand, the vertical green dashed-dotted line indicates the dimer resonance, $X_1 + X_C \simeq 0$, which takes place at longer wavelengths $\lambda = 1564\,\mathrm{nm}$. The backscattering spectra of both the single nanoparticle and the nanoparticle dimer are depicted in figure 2(b), where backscattering peaks appear at the single nanoparticle and the nanoparticle dimer resonances, respectively. In addition, the backscattering cross-section peak of the nanoparticle dimer is 3.04, i.e., approximately 4.24 times larger than the backscattering bound (1) for the dipolar response $\sigma_b(\widehat{\mathbf{k}}_i) = (1.5)^2/\pi \simeq 0.72$.

This example confirms that dimers composed of identical nanoparticles actually are excellent backscatterers. However, this 3.04 value is still below the theoretical bound (9). Consequently, the dimer should be optimized further. In this regard, the circuit model approach, i.e., equations (17), (18), suggests that the backscattering response of the dimer can be enhanced by shifting the resonance of the first particle towards higher frequencies. Fortunately, the resonance frequency of Al:ZnO nanoparticles can be tuned by adjusting the plasma frequency through the doping level. Therefore, the dimer can be optimized by assigning to the first particle a plasma frequency $\omega_{p1} = \omega_p(1 + \Delta\omega_p)$, larger than the plasma frequency of the second particle $\omega_p$. The colormap in figure 3(a) represents the backscattering cross-section peak as a function of the inter-particle separation, $d$, and the plasma frequency deviation $\Delta\omega_p$. Being consistent with the circuit model, the backscattering cross-section peak at each separation distance is optimized by a different plasma frequency deviation $\Delta\omega_p$ indicated by the dashed black line. In addition, this optimal backscattering peak approaches the bound (9) as the inter-particle separation tends to zero.

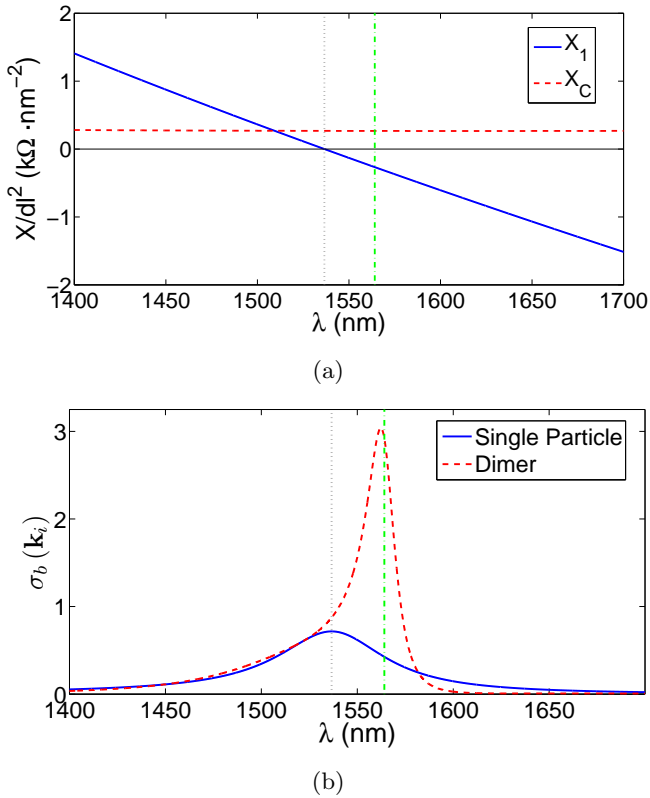

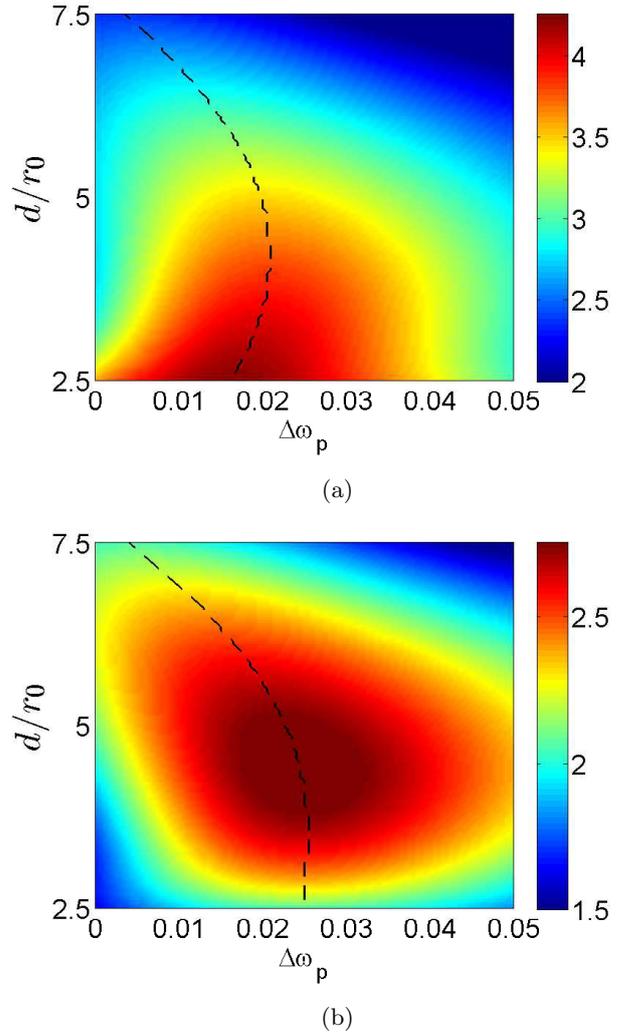

FIG. 2. (a) Particle, $X_1$, and coupling, $X_C$, reactances of a nanoparticle dimer composed of identical lossless plasmonic particles of $r_0 = 100$ nm radius, separated by the distance $d = 3r_0$. (b) Backscattering cross-section of the single plasmonic particle and the nanoparticle dimer. Vertical black dotted line indicates the single particle resonance frequency $X_1 \simeq 0$. Vertical green dashed dotted line indicates the dimer resonance: $X_1 + X_C \simeq 0$.

FIG. 3. Colormap of the backscattering cross-section peak of (a) lossless ($\omega_c = 0$) and (b) lossy ($\omega_c = 2\omega_p \times 10^{-3}$) plasmonic nanoparticle dimers, as a function of inter-particle separation $d$ and plasma frequency deviation $\Delta\omega_p$. The dashed black line indicates the position of the backscattering maximum for each separation distance.

Although our methodology identifies the optimal configurations that reach the theoretical bounds in the ideal lossless case, the dissipation losses cannot be neglected in the design of plasmonic nanoparticle architectures. For instance, increasingly smaller separation distances between the particles are inevitably associated with higher Q resonances, and, consequently, a higher sensitivity against losses. Therefore, in practice, the backscattering cross-section is not optimized for the smallest separation distance. In order to illustrate this fact, figure 3(b) depicts the backscattering cross-section peak as a function of the inter-particle separation, $d$, and the plasma frequency deviation $\Delta\omega_p$, for realistic losses $\omega_c = 2\omega_p \times 10^{-3}$. In contrast to the lossless case, the backscattering cross-section peak is optimized at the specific separation distance $d = 4.5\,r_0$, and plasma frequency deviation $\Delta\omega_p = 0.025$. At this optimal point, the backscattering cross-section approximately equals 2.8. Although this value is smaller than the upper bound (9), it nevertheless represent a 3.9 enhancement factor with respect to the individual nanoparticle bound. Therefore, it can be concluded the superbackscattering nanoparticle dimers can be successfully designed with realistic plasmonic nanoparticles.

In order to validate our methodology, figure 4(a) depicts a comparison of the backscattering spectra obtained at the optimal point ($d = 4.5\, r_0, \Delta\omega_p = 0.025$) with the circuit model (CM) methodology, and with COMSOL Multiphysics 5.0 (Num.). It is apparent from the figure that there is an excellent agreement between both methods, validating the circuit model approach. In order to complete the description of the dimer superbackscattering behavior, figure 4(b) depicts the electric field at the dimer resonance ($\lambda = 1575$ nm). Note that the particles appear cut in half due to the symmetry planes employed in the numerical simulation. The observed electric field distribution confirms the fact that the optimal backscattering response is based on the excitation of electric dipoles of similar magnitude but opposite directions. Moreover, figure 4(c) represents the scattering directivity pattern in the XZ- (solid line) and XY- (dashed line) planes at the dimer resonance ($\lambda = 1575$ nm). As noted by the bound (1), the superbackscattering response is associated with balanced forward-backward superdirective patterns, $D_{\text{scat}}(-\widehat{\mathbf{k}}_i) \simeq D_{\text{scat}}(\widehat{\mathbf{k}}_i) \simeq 3.4$. This result illustrates the fact that the superbackscattering response is constructed by the simultaneous action of the extraction (forward scattering) and re-radiation backward scattering mechanisms.

## V. HIGH PERMITTIVITY NANOPARTICLE DIMERS

Pairs of plamonic nanoparticles are but one of the many implementations of superbackscattering dimers. As a matter of fact, the dimers described in this work can be composed of any subwavelength particle exhibiting a strong dipolar resonance. As an example of this generality, this section addresses the design of superbackscattering dimers based on the magnetic dipole resonances occurring in high permittivity particles. Specifically, we make use of particles made of Tellurium (Te). The choice of Te is motivated by its high refraction index (e.g., $n_{Te} = n'_{Te} - jn''_{Te}$, with $n'_{Te} \sim 5.56$ at $10\,\mu$m [55]), and its very low losses in the thermal infrared. In fact, spectroscopy studies reveal that Te losses might be as small as $n''_{Te} \simeq 0.0004$ at $10\,\mu$m [55]. However, experiments involving the scattering from particles fit to losses that are two orders of magnitude larger: $n''_{Te} \simeq 0.04$ at $10\,\mu$m [56]. Thus, in order to have a complete perspective on the impact of losses in Te particles, low $n''_{Te} \simeq 0.0004$, high $n''_{Te} \simeq 0.04$, and even intermediate $n''_{Te} \simeq 0.004$ losses are studied here.

The design of the Te particle dimer can be carried out by using exactly the same procedure adopted for the plasmonic nanoparticle dimers. In this manner, the design starts with a dimer composed of identical particles. Even the same circuit model approach can be used, provided that the particles self-impedance is changed to $Z_p = -\eta_0(k_0 dl)^2 / \left(6\pi\, b_{11}^{oTE}\right)$, where $b_{11}^{oTE}$ is the scattered field coefficient of the particle's magnetic dipole ($TE^{o11}$) mode [40]. In this manner, the design starts with a dimer composed of identical particles. The radius of each particle is initially set at $r_1 = r_2 = 0.88\,\mu$m, so that the mag-netic dipole resonance of each individual particle takes place at $\lambda \simeq 10\mu m$. Next, the backscattering response is refined by increasing the resonance frequency of the first particle. In this case, this is simply accomplished by reducing its size, i.e., $r_1 = s \times r_2$, where $s$ is a scaling factor to be optimized.

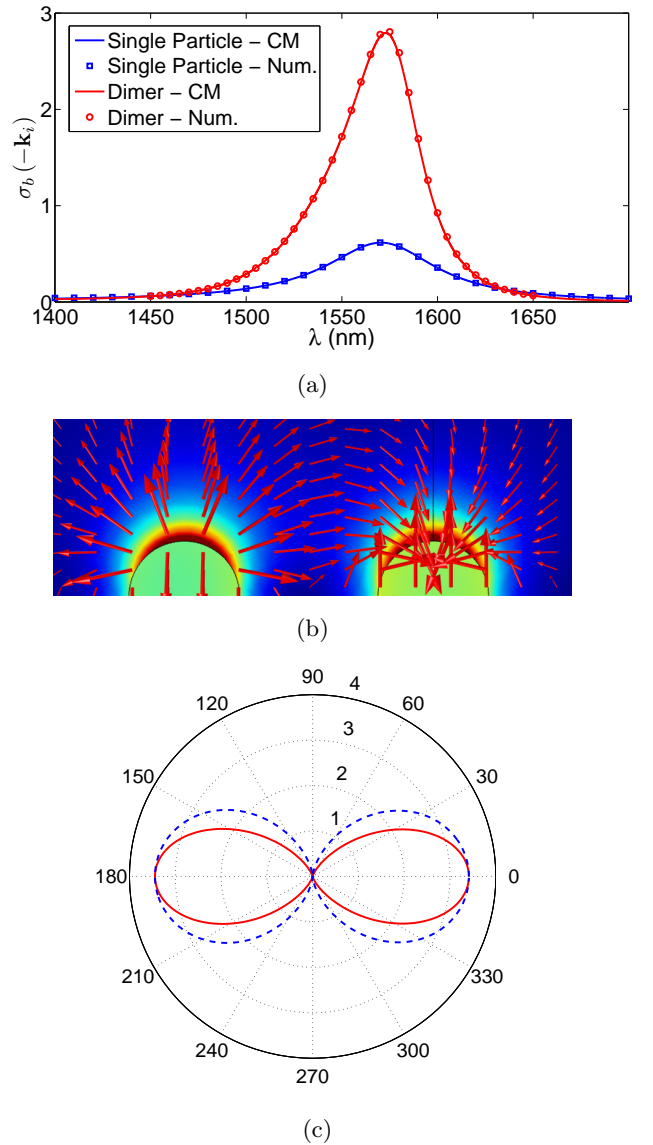

(a)

(b)

(c)

FIG. 4. (a) Backscattering cross-section spectra of the optimized plasmonic dimer and the individual nanoparticle. Comparison of the circuit model (CM) and COMSOL (Num.) predictions. (b) Electric field distribution in the XZ plane at the dimer resonance ($\lambda = 1575$ nm). The particles appear cut in half, which is due to the symmetry planes employed in the numerical simulation. (c) Scattering directivity pattern in the XZ- (solid line) and XY- (dashed line) planes at the dimer resonance ($\lambda = 1575$ nm).

Figure 5 depicts the backscattering cross-section peak as a function of the inter-particle separation, $d$, and scaling factor $s$. The figure includes the results of (a) low $n''_{Te} \simeq 0.0004$, (b) intermediate $n''_{Te} \simeq 0.004$, and (c) high $n''_{Te} \simeq 0.04$ losses. In the low loss case (see figure 5(a)), the backscattering cross-section approaches the 4.5 value of the bound (9) as the separation distance decreases. In contrast, the backscattering cross-section is optimized at specific pairs of the separation distance and scaling factor for the intermediate and high losses (see figures 5(b) and 5(c), respectively). In particular, the optimal pairs are $d = 3.25\,r_0$ and $s = 0.987$ for intermediate losses, and $d = 4.5\,r_0$ and $s = 0.985$ for high losses. In general, the larger the losses, the larger the optimal separation. Naturally, the maximal backscattering cross-section decreases along with losses, and the intermediate and high loss cases feature global maxima of 3.05 and 1.2, respectively. These values correspond to enhancement factors of 4.24 and 1.67, respectively. Therefore, our analysis suggests that, even in the presence of moderate to high losses, superbackscattering dimers composed of high permittivity particles can be successfully implemented. This conclusion can be extrapolated to any frequency regime in which high permittivity materials are available.

Again, the design procedure is validated with the use of numerical simulations. To this end, figure 6(a) depicts a comparison between the predictions on the backscattering spectra of the optimized Te-particle dimer with intermediate losses, obtained with the circuit model (CM) and COMSOL Multiphysics 5.0 (Num.). The spectra of the individual particle is also included for the sake of comparison. The figure shows that there is a good agreement between both approaches, validating the use of the CM. However, the agreement is not as excellent as it was for the plasmonic dimer (see figure 4(a)). The reason behind this slightly larger disagreement is the excitation of electric dipoles in the individual particles. Note that the CM only takes into account the magnetic dipole mode excited in the Te particles, and neglects the electric dipole response, not necessarily small in a high permittivity particle. While figure 6(a) reveals that this is actually a good approximation near the magnetic dipole resonance, the accuracy of the CM is even better for particles whose response is dominated by the electric dipole (e.g., the plasmonic particles presented in 4(a)). This outcome occurs because of the scattering responses omitted by the circuit model (i.e., the magnetic dipole and HOMs excited by the plasmonic particles) are significantly smaller. Consequently, the predictions of the CM are reasonably accurate in all the presented cases, enabling the intuitive and computationally inexpensive design of superbackscattering dimers.

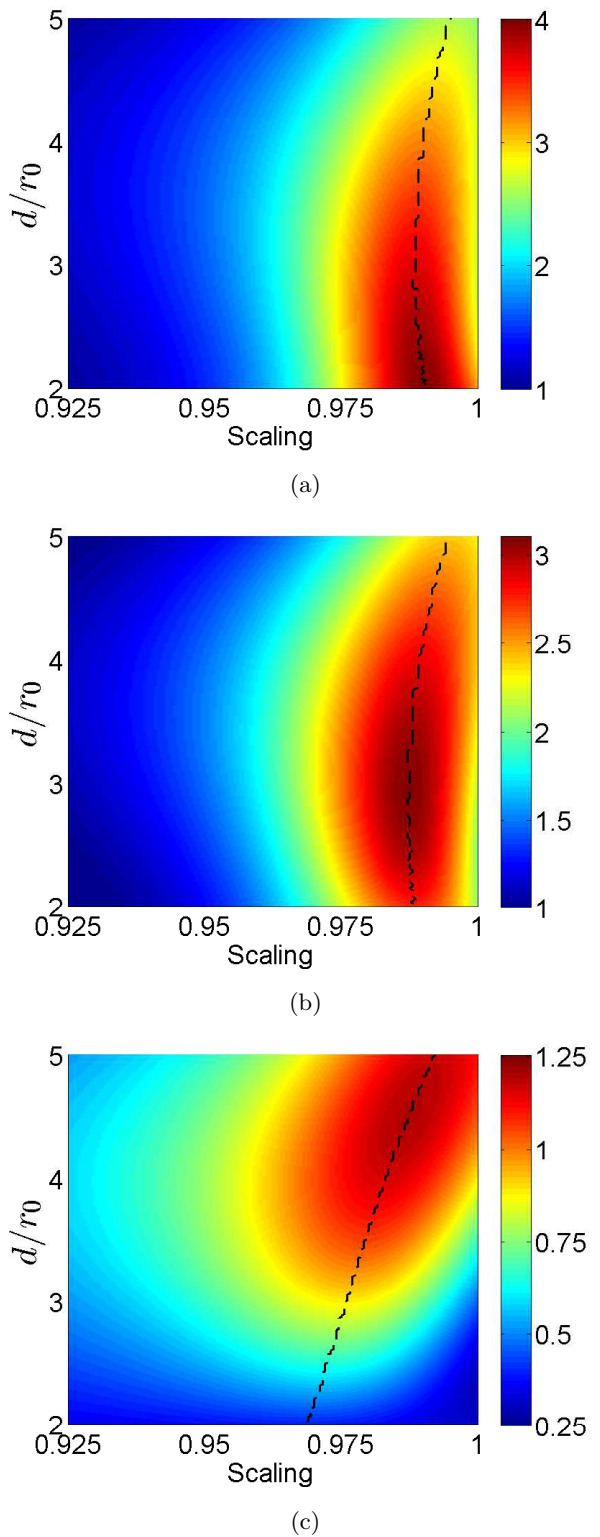

FIG. 5. Colormap of the backscattering cross-section peak of the Te particle dimers, as a function of the particle separation $d$ and the scaling factor $s$ of the first element. (a) Low $n''_{Te} \simeq 0.0004$, (b) intermediate $n''_{Te} \simeq 0.004$, and (c) high $n''_{Te} \simeq 0.04$ losses. The dashed black line indicates the position of the backscattering maximum for each separation distance.

To shed yet more light onto the polarization processes excited in the dimer, figure 6(b) represents the electric field distribution at the dimer resonance ($\lambda = 10.03\,\mu$m). In this case, the rotational character of the fields evidences the excitation of out of phase magnetic dipole moments. To finalize the description, the scattering directivity patterns at the XZ- (solid line) and XY- (dashed line) planes at the dimer resonance ($\lambda = 10.03\,\mu$m) are depicted in figure 6(c). Again, the superbackscattering response is accompanied by superdirective scattering pointing to both the forward ($D_{\rm scat}(\widehat{\bf k}_i) \simeq 3.1$) and the backward ($D_{\rm scat}(-\widehat{\bf k}_i) \simeq 3.3$) directions.

## VI. CONCLUSIONS

This work has addressed the design of superbackscattering dimers. We analytically derived the conditions under which the backscattering cross-section of a nanoparticle dimer is maximized. Interestingly, the largest backscattering response is observed exactly when the dipole excited in the particles composing the dimer are of the same magnitude but opposite directions. When this condition is met, nanoparticle dimers can theoretically feature a backscattering cross-section 6.25 times larger than that of the individual particles. Moreover, it was demonstrated that nanoparticle dimers approaching this enhancement factor can be straightforwardly designed by using a simple circuit model. Naturally, the performance of these systems will be limited in practice by dissipation damping and other practical issues. Despite this fact, our examples demonstrate that, even when realistic losses are taken into account, both plasmonic and high-permittivity particle dimers can exhibit fourfold enhancements of the backscattering cross-section with respect to the individual particles. These results encourage us to believe that the development of superbackscattering nanparticles dimers is possible. Due to the complementary interest of backscattering measurements, we are confident that these nanoparticle architectures will be of great interest for variety of spectroscopy, communication, remote sensing, manipulation and/or imaging systems.

## Appendix A: Derivation of the optimal excitation coefficients

This appendix summarizes the maximization process that identifies the optimal excitation coefficients and upper bounds of the backscattering and forward scattering cross-sections of electrically small nanoparticle dimers. To this end, and in contrast with figure 1(a), here we assume that the particles composing the dimer are located along the x-axis, and the incident field is $\widehat{\bf z}$-linearly polarized. This new coordinate system is particularly convenient because the scattered electric field in the far zone is $\widehat{\boldsymbol{\theta}}$ polarized. Specifically, the scattering pattern of the

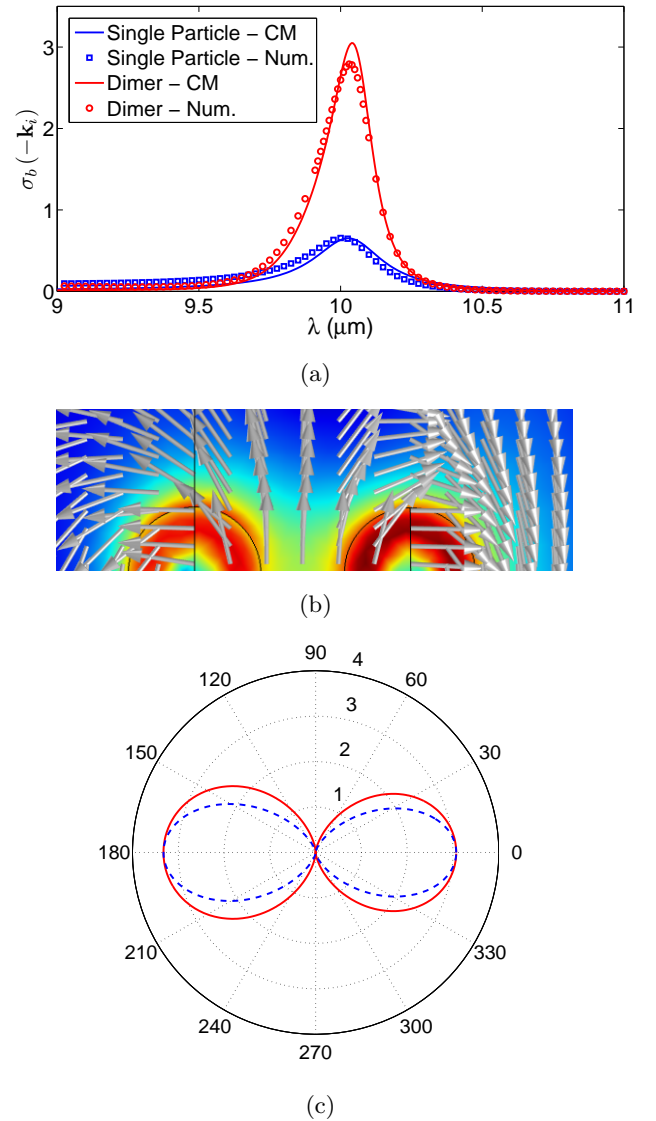

FIG. 6. (a) Backscattering cross-section spectrum of the optimal ($d = 3.25\,r_0$, $s = 0.987$) Te-particle dimer with intermediate losses ($n_{Te} = 5.56 - j0.004$). Comparison between the predictions of the circuit model (CM) and COMSOL (Num.). (b) Electric field distribution in the XZ plane at the dimer resonance ($\lambda = 10.03\,\mu$m). The particles appear cut in half due to the symmetry planes employed in the numerical simulation. (c) Scattering directivity pattern in the XZ- (solid line) and XY- (dashed line) planes at the dimer resonace.

dimer in this new coordinate system is given by

$$\mathbf{f}\left(\widehat{\mathbf{r}}\right) = \widehat{\boldsymbol{\theta}}\sin\theta \left( A_1 e^{-j\frac{k_0 d}{2}\widehat{\mathbf{r}}\cdot\widehat{\mathbf{x}}} + A_2 e^{j\frac{k_0 d}{2}\widehat{\mathbf{r}}\cdot\widehat{\mathbf{x}}} \right) \quad \text{(A1)}$$

Note also that the scattering directivity (3) can be written alternatively as: [44, 46]

$$D_{\rm scat}\left(\widehat{\mathbf{r}}\right) = \frac{3}{2}\sin^2\theta \frac{\left|A_1 + A_2 e^{jk_0 d\widehat{\mathbf{r}}\cdot\widehat{\mathbf{x}}}\right|^2}{\left|A_1\right|^2 + \left|A_2\right|^2 + 3\,H\,{\rm Re}\left\{A_1 A_2^*\right\}} \quad \text{(A2)}$$

with $H = H(k_0 d)$ defined as

$$H(k_0 d) = \frac{\sin(k_0 d)}{k_0 d}\left[1 - \frac{1}{(k_0 d)^2}\right] + \frac{\cos(k_0 d)}{(k_0 d)^2} \quad (A3)$$

Next, in order to study electrically small dimers we observe the limit

$$\lim_{k_0 d \to 0} H(k_0 d) \simeq \frac{2}{3}\left[1 - \frac{(k_0 d)^2}{5}\right] \quad (A4)$$

In this manner, the scattering directivity of an electrically small ($k_0 d \to 0$) dimer approximately reduces to

$$D_{\text{scat}}(\widehat{\mathbf{r}}) \simeq \frac{3}{2}\sin^2\theta \frac{|A_1 + A_2(1 + jk_0 d\widehat{\mathbf{r}}\cdot\widehat{\mathbf{x}})|^2}{|A_1 + A_2|^2 - \frac{2}{5}(k_0 d)^2 \operatorname{Re}\{A_1 A_2^*\}} \quad (A5)$$

If we take the $k_0 d \to 0$ limit in (A5) with constant coefficients, then the scattering directivity reduces to a dipolar pattern $D_{\text{scat}}(\widehat{\mathbf{r}}) \simeq (3/2)\sin^2\theta$. Therefore, in order to get a response different from a dipole pattern from an electrically small dimer, the excitation coefficients must increase as the separation distance decreases. The coefficients that satisfy this condition, while keeping a finite scattering pattern, can in general be written as:

$$A_1 = -\frac{a_0}{k_0 d}, \quad A_2 = a_0\left(\frac{1}{k_0 d} + a_1\right) \quad (A6)$$

By using these coefficients, the scattering directivity of an electrically small dimer is given by

$$D_{\text{scat}}(\widehat{\mathbf{r}}) \simeq \frac{3}{2}\sin^2\theta \frac{|a_1 + j\widehat{\mathbf{r}}\cdot\widehat{\mathbf{x}}|^2}{|a_1|^2 + \frac{2}{5}} \quad (A7)$$

This equation can be maximized at any specific direction by using standard techniques. For example, it is clear from (A7) that the forward scattering directivity, $\widehat{\mathbf{r}}\cdot\widehat{\mathbf{x}} = 1$, is maximized with an imaginary coefficient $a_1 = jc_1$. Maximization with respect to $c_1$ leads to the solution $c_1 = 2/5$. By using this value, it is found that the forward scattering directivity of an electrically small nanoparticle dimer is upper bounded by

$$D_{\text{scat}}(\widehat{\mathbf{x}}) \leq \frac{21}{4} = 5.25 \quad (A8)$$

This analytical result is consistent with the numerical optimizations carried out in [45, 46].

On the other hand, the forward-backward directivity product can be written as:

$$D_{\text{scat}}(\widehat{\mathbf{x}})D_{\text{scat}}(-\widehat{\mathbf{x}}) = \frac{9}{4}\frac{\left(|a_1|^2 + 1\right)^2 - 4\operatorname{Im}\{a_1\}^2}{\left(|a_1|^2 + \frac{2}{5}\right)^2} \quad (A9)$$

It is apparent from (A9) that the forward-backward directivity product is maximized for $a_1$ being a real number. The maximization for such a real number results in the optimal value $a_1 = 0$, i.e., the forward-backward directivity product is maximized when the excitation coefficients satisfy $A_2 = -A_1 = a_0/k_0 d$. This condition physically corresponds to the excitation of dipoles with the same magnitude but opposite directions. At this maximal point we find the upper bound of the forward-backward directivity product of an electrically small nanoparticle dimer to be:

$$D_{\text{scat}}(\widehat{\mathbf{x}})D_{\text{scat}}(-\widehat{\mathbf{x}}) \leq \left(\frac{3\cdot 5}{4}\right)^2 = 14.06 \quad (A10)$$

Consequently, the upper bound of the backscattering cross-section of an electrically small nanoparticle dimer can be found by introducing (A10) into (1). It is:

$$\sigma_b\left(-\widehat{\mathbf{k}}_i\right) \leq \frac{1}{\pi}\left(\frac{3\cdot 5}{4}\right)^2 = 4.48 \quad (A11)$$

---